\documentclass[11pt,preprint,longabstract]{aastex}
\shorttitle{Calcium Emission in Interacting Binary Be Stars}
\shortauthors{Pavel Koubsk\'y et al.}

\begin{document}

\title{\Large Infrared Calcium Triplet Emission in Interacting Binary Be Stars}

\author{Pavel Koubsk\'y\altaffilmark{1}, 
  Lenka~Kotkov\'a$^1$, 
  Viktor~Votruba$^1,3$,
  Mirek~\v Slechta$^1$, 
  Petr~\v Skoda$^1$,
  Petr~Harmanec$^2$,
  Jana~Nemravov\'a$^2$,
  Kl\'ara~\v Sejnov\'a$^1,3$, 
  \v S\'arka~Dvo\v{r}\'akov\'a$^3$, and 
  Daniela~Kor\v c\'akov\'a$^2$
} 

\affil{
$^1${\rm Astronomical Institute, Academy of Sciences of CR, Fri\v{c}ova 298,  CZ~251\,65 Ond\v rejov,
\\ Czech Republic}
\\email: {\tt koubsky(lenka,slechta,skoda,klara)@sunstel.asu.cas.cz, votruba@physics.muni.cz } 
\\
$^2${\rm Astronomical Institute of the Charles University, Faculty of Mathematics and Physics, \\ V~Hole\v{s}ovi\v{c}k\'{a}ch 2, CZ~180\,00 Praha 8, Czech Republic} 
\\email: {\tt hec(kor)@sunstel.asu.cas.cz, janicka.ari@seznam.cz}
\\
$^3$ {\rm Department of Theoretical Physics and Astrophysics, Masaryk University, Kotl\'{a}\v{r}sk\'a  2, CZ~611\,37~Brno, Czech Republic  }
\\email: {\tt dvorakova@sunstel.asu.cas.cz}
}
\begin{abstract}
Polidan (1976) suggested that Be stars showing the CaII IR triplet in emission are interacting binaries.
 With the advent of the Gaia satellite, which will host a spectrometer to observe stars in the range 8470--8750 \AA, we carried out a spectroscopic survey of 150 Be stars, including  Be binaries.  We show that the Ca II triplet in emission, often connected with emission in Paschen lines, is an indicator of a peculiar
 environment in a Be star disc rather than a signature of an interacting binary Be star. However, Ca II emission without visible emission in Paschen lines is observed in interacting binary stars, as well as in  peculiar objects. During the survey, a new interacting Be binary - HD 81357 - was discovered. 
\keywords{stars: emission-line, Be, binaries: spectroscopic, (stars:) circumstellar matter, stars: individual (HD 81357)}
\end{abstract}

\section{Introduction}

An atlas of O-G0 stars in the infrared spectral region was presented by Andrillat et al. (1995), but no systematic attempt has been undertaken so far to investigate the spectra of Be stars in the RVS domain of Gaia. Therefore, we decided to carry out a large scale documentation project on Be stars. A very brief description of the Ca~II IR lines in Be stars can be found in Jaschek et al. (1988).  The aim of the project was to compare H$\alpha$ and Paschen spectra of Be stars and exploit the possibility to derive the properties of H$\alpha$ from the observations of higher members of the Paschen series, as Gaia instrumentation will complement the medium resolution IR spectra by very poor spectral resolution for H$\alpha$ line. In addition, one can study the behaviour of the calcium triplet in early-type emission stars and interacting binaries.  For the survey, we used the Ond\v{r}ejov 2-m telescope equipped with a Coud\'{e} spectrograph. Pairs of spectra covering H$\alpha$ and Gaia RVS region have been so far obtained for about 150 Be stars.
\section{Data collection and reduction }
The spectra were secured with the Ond\v rejov single order spectrograph + SITe 2000x800 chip (R=12700, 6250-6750 and 8392-8900 \AA). The stellar and calibration spectra (Th-Ar and tungsten lamps) were reduced using IRAF.

\section{Results}
{\bf Calcium triplet emission in interacting binary Be stars.} The Ca II triplet lines coincide with Paschen P16, 15, and 13. For a strong emission object like $\chi$ Oph, the contribution of calcium to the Paschen emission is marginal, if any (no difference in intensity between P14 and P15/P16). However, in ``normal" Be stars (e.g. V731~Tau or $\gamma$~Cas) the calcium emission is growing with the decreasing strength of H$\alpha$ emission. It is very pronounced in interacting binaries AX~Mon, KX~And, BR~CMi, but not visible in systems like RX~Cas, CX~Dra or even $\beta$ Lyr.  Spectra of binaries HR~2142, 59~Cyg or $\phi$ Per mimic rather ``normal" Be stars in the infrared, while $\psi$~Per resembles KX~And. Thus, after 35 years, the suggested connection between the behaviour of calcium emission and interacting binaries is still open.

\begin{figure}[h]
\begin{center}
 \includegraphics[width=3.8in,height=1.82in]{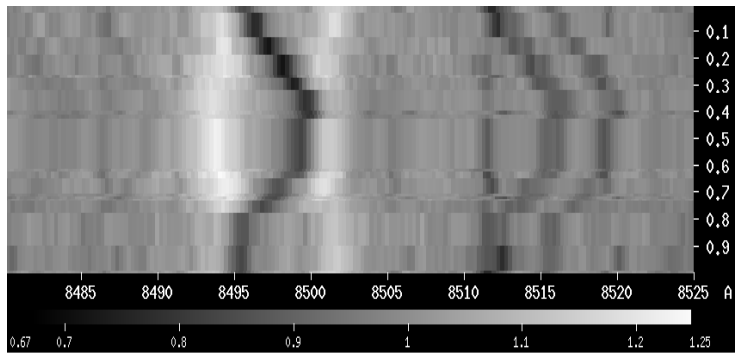} 
 \caption{Trailed spectra of HD~81357 versus the phase of the orbital period $P$=$33.\!\!^{\rm d}73$. The radial-velocity curve of the secondary is clearly defined by the motion of the Ca II 8498 \AA\ line which shows an emission component too. The emission is moving in anti-phase (possible disc around the primary?). The spectra were secured at Ond\v rejov during March 2011 - April 2012.}
   \label{fig1}
   \end{center}
\end{figure}

{\bf HD 81357  new (interacting?) Be binary.} In the course of our survey of Be stars in H$\alpha$ and RVS regions, an interesting object -- HD 81357  was discovered. Its red and infrared spectrum closely resembles BR~CMi, an interacting Be binary (Royer et al., 2007). We derived the first orbital elements: $P$=$33.\!\!^{\rm d}73$, $K_{2}$=79 km~s$^{-1}$, $e$=0. Trailed spectra are shown in Figure 1. The orbital period applied on Hipparcos photometric data gives double-wave curve, while the 25.7-day period suggested by Koen and Eyer (2002) yields simple curve.  The semi-amplitude of the secondary (cool) star is based on the measurements of Fe I and Ca II absorption lines. As in case of BR~CMi, the lines of primary component are not readily seen. The estimated semi-amplitude of the primary (hot) star is in the range of  5--15 km~s$^{-1}$ (from the motion of H$\alpha$ and Ca II emission).

{\it Aknowledgements.} This research was supported by grants ESA PECS 98058, GA\v CR P209/10/0715,  P205/09/P476, and by the agreement between SAV and AVCR. These results were presented at IAU Symposium 282 ``From Interacting Binaries to Exoplanets: Essential Modelling Tools'' in Slovakia. We aknowledge the use of one spectrum kindly secured by Dr. A. Aret.

{}

\end{document}